\documentclass[11pt,a4paper]{article} \usepackage{amsbsy,epsfig,cite}

\def\Gambb               {\ensuremath {\Gamma_{\bb}}}
\def\Gamhad              {\ensuremath {\Gamma_{\mathrm{had}}}}
\newcommand{\Rb}         {\ensuremath{R_{b}}}
\newcommand{\Rudsc}      {\ensuremath{R_{udsc}}}
\newcommand{\chisq}      {\ensuremath{\chi^2}}
\newcommand{\mc}         {Monte Carlo}

\newcommand{\mhic}       {multihadronic}
\newcommand{\mh}         {multihadron}
\newcommand{\Nmh}        {\ifmmode {N_{mh}}\else    {$N_{mh}$}\fi}
\newcommand{\bbaryon}    {b--baryon}
\newcommand{\bmeson}     {B meson}
\newcommand{\bhadron}    {b-hadron}
\newcommand{\dEdx}       {\ensuremath {{\mathrm dE}/{\mathrm dx}}}
\newcommand{\Pt}         {\ensuremath {p_t}}
\newcommand{\mom}        {\ensuremath {p}}
\newcommand{\pz}         {\ensuremath {p_z}}
\newcommand{\MeVcc}      {\ensuremath {\mathrm{MeV/c}^{2}}}
\newcommand{\GeVc}       {\ensuremath {\mathrm{GeV/c}}}
\newcommand{\GeVcc}      {\ensuremath {\mathrm{GeV/c^2}}}
\newcommand{\Bd}         {\ensuremath {\mathrm{B^0}}}
\newcommand{\Bs}         {\ensuremath {\mathrm{B^0_s}}}
\newcommand{\Lamb}       {\ensuremath {\Lambda}}
\newcommand{\pion}       {\ensuremath {\pi}}
\newcommand{\proton}     {\ensuremath {\mathrm{p}}}
\newcommand{\Clam}       {\ensuremath {\Lambda_{\mathrm{c}}}}

\newcommand{\Blam}       {\ensuremath {\Lambda_{\mathrm{b}}}}
\newcommand{\LambPPi}    {\ensuremath {\Lamb \rightarrow \proton\pion}}
\newcommand{\BlamLambX}  {\ensuremath {\Blam\rightarrow\Lambda\X}}
\newcommand{\BlamLambLepX}{\ensuremath{\Blam\rightarrow\Lambda\ell\X}}
\newcommand{\BLambX}     {\ensuremath {\mathrm{B}\rightarrow\Lambda\X}}
\newcommand{\FragLambX}  {\ensuremath {\mathrm{frag}\rightarrow\Lambda\X}}
\newcommand{\Vcb}        {\ensuremath {\mathrm{V_{cb}}}}
\newcommand{\fbBlam}     {\ensuremath {f(\mathrm{b}\rightarrow\Blam)}}
\newcommand{\epsLam}     {\ensuremath {\epsilon_{\Lamb}}}
\newcommand{\epsBlamLambX}{\ensuremath {\epsilon^{b}_{\BlamLambX}}}
\newcommand{\epsdv}      {\ensuremath {\epsilon_{sig}^b}}
\newcommand{\epsb}       {\ensuremath{\epsilon_{\mathrm{b}}}}
\newcommand{\epudsc}     {\ensuremath{\epsilon_{\mathrm{udsc}}}}
\newcommand{\Zzero}      {\ensuremath {\mathrm{Z}^{0}}}

\newcommand{\bb}         {\ensuremath {\mathrm{b\bar{b}}}}
\newcommand{\qq}         {\ensuremath {\mathrm{q\bar{q}}}}
\newcommand{\Ztoqq}      {\ensuremath {\Zzero\rightarrow\qq}}
\newcommand{\Ztobb}      {\ensuremath {\Zzero\rightarrow\bb}}
\newcommand{\Dplus}      {\ensuremath {\mathrm {D^+}}}
\newcommand{\X}          {\ensuremath {\mathrm X}}

\newcommand{\PBRBlamLambX}   {\ensuremath{f(\mathrm{b}\rightarrow\Blam)\cdot
  \mathrm{BR}(\BlamLambX)}}
\newcommand{\PBRBlamLambLepX}{\ensuremath{f(\mathrm{b}\rightarrow
  \Blam)\cdot \mathrm{BR}(\BlamLambLepX)}}

\newcommand{\BRLambPPi}      {\ensuremath {\mathrm{BR}(\Lamb\rightarrow
  \proton\pion)}}
\newcommand{\GamBlamLambX}   {\ensuremath {\Gamma(\BlamLambX)}}
\newcommand{\GamBlamLambLepX}{\ensuremath {\Gamma(\BlamLambLepX)}}
\newcommand{\BRBlamLambX}   {\ensuremath {\mathrm{BR}(\BlamLambX)}}
\newcommand{\BRBlamLambLepX}{\ensuremath {\mathrm{BR}(\BlamLambLepX)}}

\newcommand{\RatGamBlamLambLepXSLASH}{\ensuremath
  {\GamBlamLambLepX/\GamBlamLambX}}
\newcommand{\RatBRBlamLambLepXSLASH}{\ensuremath
  {\BRBlamLambLepX/\BRBlamLambX}}
\newcommand{\RatBRBlamLambLepX}{\ensuremath{\RatBRBlamLambLepXSLASH}}
\newcommand{\RatGamBlamLambLepX}{\ensuremath{\RatGamBlamLambLepXSLASH}}

\newcommand{\PPEnum}          {CERN-EP/98-186}

\newcommand{\prdate}          {25 November 1998}
\newcommand{\symtitle}[1]{{$\boldsymbol{\protect #1}$}}
\newcommand{\etal}{{\it et~al.}}

\newcommand{\NumTkmhStatus}   {3\,554\,212}
\newcommand{\NumTkmhAllCuts}  {2\,323\,302}
\newcommand{\Pmincut}         {5}
\newcommand{\NumLambDvCut}    {1582}
\newcommand{\BlamPercent}     {($37.4\pm 5.3$)\%} 
\newcommand{\NumBlamCorr}     {\ensuremath{592\pm 83}}
\newcommand{\Lambeff}         {\ensuremath{(11.7\pm 0.6)\%}}
\newcommand{\PBRAnswer}       {\ensuremath {(2.67\pm 0.38 (stat) ^{+0.67}
                               _{-0.60}(sys))\%}}
\newcommand{\OPALPBR}         {\ensuremath {(3.50\pm 0.32(stat)\pm 0.35(sys))\%}}
\newcommand{\BRAnswer}        {\ensuremath{(35 ^{+14}_{-12})\%}}
\newcommand{\btobbaryon}      {\ensuremath{(10.1 ^{+3.9}_{-3.1})\%}}
\newcommand{\btobmeson}       {\ensuremath{(89.9 ^{+3.1}_{-3.5})\%}}

\newcommand{\OPALlamlep}        {bib-OPALpr055}
\newcommand{\OPALleptonID}      {bib-OPALpr146}
\newcommand{\DELPHIpbr}         {bib-DELPHI95}
\newcommand{\TauLifeUpdated}    {bib-OPALpr112}
\newcommand{\TauLife}           {bib-OPALpr072}
\newcommand{\OPALnewRb}         {bib-OPALpr257}

\newcommand{\OPALPGagnon}       {bib-OPALpr189}
\newcommand{\OPALdetector}      {bib-OPALpr021}

\newcommand{\OPALSI}            {bib-OPALip017}

\newcommand{\OPALnewdedx}       {bib-OPALdedx}

\newcommand{\OPALblampol}       {bib-OPALpr255}
\newcommand{\OPALblamlife}      {bib-OPALpr138}
\newcommand{\OPALlambdaID}      {bib-OPALpr061}
\newcommand{\OPALblamleppbr}    {bib-OPALpr138}
\newcommand{\OPALgammabb}       {bib-OPALpr107}
\newcommand{\OPALcone}          {bib-OPALpr097}
\newcommand{\OPALbstars}        {bib-OPALpr116}

\newcommand{\ALEPHblamlep}      {bib-ALEPHblamlep}

\newcommand{\BarlowBeeston}     {bib-BarlowBeeston-fitter}
\newcommand{\CLEOBLamb}         {bib-CLEOBLamb92}

\newcommand{\JETSET}            {bib-JETSET74}
\newcommand{\PetersonFrag}      {bib-PetersonFrag}

\newcommand{\PDG}               {bib-PDG1998}

\begin{document}
\bibliographystyle{plain}


\begin{titlepage}
\def\toprule{\noalign{\hrule \medskip}}
\def\midrule{\noalign{\medskip\hrule }}
\def\botrule{\noalign{\medskip\hrule }}
\flushbottom
%
\begin{center}
   {\large\bf EUROPEAN LABORATORY FOR PARTICLE PHYSICS}
\end{center}
%
\bigskip
\begin{flushright}
  \PPEnum \\ \prdate
\end{flushright}
\bigskip\bigskip\bigskip\bigskip\bigskip
%
\begin{center}  
  {\huge\bf
    A Measurement of\\
    \vspace{1mm}
    the Product Branching Ratio\\
    \vspace{1mm} \symtitle{\PBRBlamLambX} \\ in \symtitle{\Zzero} Decays}
\end{center}
\bigskip\bigskip
%
\begin{center}
  {\LARGE The OPAL Collaboration}
\end{center}
\bigskip\bigskip\bigskip
%
\begin{center}
\begin{abstract}
\noindent The product branching ratio, \PBRBlamLambX , where \Blam\
denotes any weakly-decaying \bbaryon, has been measured using the OPAL
detector at LEP. \Blam's are selected by the presence of energetic
\Lamb\ particles in bottom events tagged by the presence of displaced
secondary vertices.  A fit to the momenta of the \Lamb\ particles
separates signal from \bmeson\ and fragmentation backgrounds.  The
measured product branching ratio is \[ \PBRBlamLambX = \PBRAnswer .\]
Combined with a previous OPAL measurement, one obtains \[
\PBRBlamLambX =\OPALPBR .\]
\end{abstract}
\end{center}
%
\bigskip\bigskip\bigskip\bigskip\bigskip\bigskip
\begin{center}
  {\large
   (Submitted to the European Physical Journal C)}
\end{center}

\end{titlepage}

\begin{center}{\Large        The OPAL Collaboration
}\end{center}\bigskip
\begin{center}{
G.\thinspace Abbiendi$^{  2}$,
K.\thinspace Ackerstaff$^{  8}$,
G.\thinspace Alexander$^{ 23}$,
J.\thinspace Allison$^{ 16}$,
N.\thinspace Altekamp$^{  5}$,
K.J.\thinspace Anderson$^{  9}$,
S.\thinspace Anderson$^{ 12}$,
S.\thinspace Arcelli$^{ 17}$,
S.\thinspace Asai$^{ 24}$,
S.F.\thinspace Ashby$^{  1}$
% ...
}\end{center}\bigskip
\bigskip
$^{  1}$School of Physics and Astronomy, University of Birmingham,
Birmingham B15 2TT, UK
\newline
$^{  2}$Dipartimento di Fisica dell' Universit\`a di Bologna and INFN,
I-40126 Bologna, Italy
\newline
$^{  3}$Physikalisches Institut, Universit\"at Bonn,
D-53115 Bonn, Germany
\newline
$^{  4}$Department of Physics, University of California,
Riverside CA 92521, USA
\newline
$^{  5}$Cavendish Laboratory, Cambridge CB3 0HE, UK
\newline
$^{  6}$Ottawa-Carleton Institute for Physics,
Department of Physics, Carleton University,
Ottawa, Ontario K1S 5B6, Canada
\newline
$^{  7}$Centre for Research in Particle Physics,
Carleton University, Ottawa, Ontario K1S 5B6, Canada
\newline
$^{  8}$CERN, European Organisation for Particle Physics,
CH-1211 Geneva 23, Switzerland
\newline
$^{  9}$Enrico Fermi Institute and Department of Physics,
University of Chicago, Chicago IL 60637, USA
\newline
$^{ 10}$Fakult\"at f\"ur Physik, Albert Ludwigs Universit\"at,
D-79104 Freiburg, Germany
\newline
$^{ 11}$Physikalisches Institut, Universit\"at
Heidelberg, D-69120 Heidelberg, Germany
\newline
$^{ 12}$Indiana University, Department of Physics,
Swain Hall West 117, Bloomington IN 47405, USA
\newline
$^{ 13}$Queen Mary and Westfield College, University of London,
London E1 4NS, UK
\newline
$^{ 14}$Technische Hochschule Aachen, III Physikalisches Institut,
Sommerfeldstrasse 26-28, D-52056 Aachen, Germany
\newline
$^{ 15}$University College London, London WC1E 6BT, UK
\newline
$^{ 16}$Department of Physics, Schuster Laboratory, The University,
Manchester M13 9PL, UK
\newline
$^{ 17}$Department of Physics, University of Maryland,
College Park, MD 20742, USA
\newline
$^{ 18}$Laboratoire de Physique Nucl\'eaire, Universit\'e de Montr\'eal,
Montr\'eal, Quebec H3C 3J7, Canada
\newline
$^{ 19}$University of Oregon, Department of Physics, Eugene
OR 97403, USA
\newline
$^{ 20}$CLRC Rutherford Appleton Laboratory, Chilton,
Didcot, Oxfordshire OX11 0QX, UK
\newline
$^{ 22}$Department of Physics, Technion-Israel Institute of
Technology, Haifa 32000, Israel
\newline
$^{ 23}$Department of Physics and Astronomy, Tel Aviv University,
Tel Aviv 69978, Israel
\newline
$^{ 24}$International Centre for Elementary Particle Physics and
Department of Physics, University of Tokyo, Tokyo 113-0033, and
Kobe University, Kobe 657-8501, Japan
\newline
$^{ 25}$Institute of Physical and Environmental Sciences,
Brunel University, Uxbridge, Middlesex UB8 3PH, UK
\newline
$^{ 26}$Particle Physics Department, Weizmann Institute of Science,
Rehovot 76100, Israel
\newline
$^{ 27}$Universit\"at Hamburg/DESY, II Institut f\"ur Experimental
Physik, Notkestrasse 85, D-22607 Hamburg, Germany
\newline
$^{ 28}$University of Victoria, Department of Physics, P O Box 3055,
Victoria BC V8W 3P6, Canada
\newline
$^{ 29}$University of British Columbia, Department of Physics,
Vancouver BC V6T 1Z1, Canada
\newline
$^{ 30}$University of Alberta,  Department of Physics,
Edmonton AB T6G 2J1, Canada
\newline
$^{ 31}$Duke University, Dept of Physics,
Durham, NC 27708-0305, USA
\newline
$^{ 32}$Research Institute for Particle and Nuclear Physics,
H-1525 Budapest, P O  Box 49, Hungary
\newline
$^{ 33}$Institute of Nuclear Research,
H-4001 Debrecen, P O  Box 51, Hungary
\newline
$^{ 34}$Ludwigs-Maximilians-Universit\"at M\"unchen,
Sektion Physik, Am Coulombwall 1, D-85748 Garching, Germany
\newline
\bigskip\newline
$^{  a}$ and at TRIUMF, Vancouver, Canada V6T 2A3
\newline
$^{  b}$ and Royal Society University Research Fellow
\newline
$^{  c}$ and Institute of Nuclear Research, Debrecen, Hungary
\newline
$^{  d}$ and Department of Experimental Physics, Lajos Kossuth
University, Debrecen, Hungary
\newline
$^{  e}$ on leave of absence from the University of Freiburg
\newline


\clearpage
\section{Introduction} \label{sec:intro}

In this paper, we present a measurement of the product branching
ratio, \PBRBlamLambX, at the \Zzero\ resonance.\footnote{ Throughout
  this paper \Blam\ refers to any weakly-decaying \bbaryon.  Charge
  conjugate modes are implied.}  In this process a b quark from
\Ztobb\ decays produces a b-flavoured baryon which decays, directly or
indirectly, into a \Lamb\ baryon and other particles.  Previous
studies of inclusive \bbaryon\ decays have emphasized semileptonic
decays of the \Blam \cite{\OPALlamlep,\OPALblamlife,\ALEPHblamlep}.

The result presented here, when combined with the semileptonic
branching ratio, \PBRBlamLambLepX, allows one to determine the ratio
\RatGamBlamLambLepX \cite{\OPALPGagnon}.  Since \GamBlamLambLepX\ 
depends on \Vcb\ and well understood leptonic currents, it may also be
possible to extract this fundamental weak parameter in a setting with
hadronic uncertainties different from those of \bmeson\ measurements
\cite{bib-bigi,bib-neubert}.

For this analysis, we select events containing a \Lamb\ particle and a
vertex significantly displaced from the \Zzero\ decay point.  This
gives a sample enriched in \Blam's.  Significant backgrounds come from
the decay of \bmeson s into \Lamb\ particles, and from \bhadron\ 
events where a high-momentum \Lamb\ is produced in the primary
hadronisation process.  These backgrounds are separated from the
signal by using a simultaneous fit to the momentum and transverse
momentum distributions of the \Lamb\ baryon.

The previously published OPAL measurement of the product branching
ratio used a ``companion baryon technique'' to identify jets
containing a \Blam \cite{\OPALPGagnon}.  That analysis used the
momenta of a \Lamb\ and an anti-baryon identified in the same
hemisphere, whereas this analysis uses the momentum and transverse
momentum of a \Lamb.  The two techniques are complementary and have
less than 20\% of events in common.

After general information about the OPAL detector and \mc\ event
simulation, we outline the event selection which provides an enriched
sample of \Blam.  The backgrounds are then addressed, followed by a
discussion of signal efficiencies.  Systematic errors are discussed in
detail. Finally, the measured value of the product branching ratio is
presented.  This is combined with the previous OPAL measurement, and
is used to update the ratio \RatBRBlamLambLepX\ from
\cite{\OPALPGagnon}.


\section{The OPAL Detector and Its Simulation} \label{sec:detector}

The OPAL detector is described in detail elsewhere
\cite{\OPALdetector}. Here, we briefly describe the components which
are particularly relevant to this analysis. Charged particle tracking
is performed by the central tracking system which is located in a
solenoidal magnetic field of 0.435~T\@.  The central tracking system
consists of a two-layer silicon micro-vertex detector \cite{\OPALSI},
a high-precision vertex drift chamber, a large-volume jet chamber and
$z$--chambers for accurately measuring track coordinates along the
beam direction.

The measurement of specific ionisation in the jet chamber, \dEdx, is
used for particle identification.  Tracks emitted at large angle to
the beam direction have up to 159 samplings providing a \dEdx\ 
resolution of 3.2\% \cite{\OPALnewdedx}.

The central detector is surrounded by a lead-glass electromagnetic
calorimeter with a wire streamer chamber as presampler. The iron
magnet yoke is instrumented with layers of streamer tubes which serve
as a hadron calorimeter and provide information for muon
identification.  Four layers of planar drift chambers surround the
hadron calorimeter and serve for tracking muons.

To obtain momentum distributions of \Lamb's from different sources,
and to evaluate efficiencies and backgrounds, we utilize 6 million
\Ztoqq\ and 3 million \Ztobb\ simulated events.  The \mc\ simulation
of the OPAL detector is described elsewhere \cite{bib-OPALip006}.  The
JETSET 7.4 string fragmentation program is used to form hadrons and
decay short-lived particles \cite{\JETSET,\OPALlambdaID}.  The
fragmentation function of Peterson \etal, is used for heavy flavors,
($\epsilon_P=0.0038$ for b quarks) \cite{\PetersonFrag,\PDG}.  For
this analysis the \Lamb\ momentum in the rest frame of \bmeson s for
\Lamb's from \bmeson\ decay is tuned to match measurements by CLEO
\cite{\CLEOBLamb}.


\section{Event Selection} \label{sec:sel}

This study uses a total of \NumTkmhStatus\ hadronic \Zzero\ decays
collected by the OPAL detector between 1991 and 1995. The method for
selecting hadronic \Zzero\ decays has been described in previous OPAL
publications \cite{\OPALnewRb,\OPALgammabb} and has an efficiency of
$(98.7\pm 0.4)\%$.
 
To select events with a clear two-jet structure, the thrust of the
event is required to be at least $0.8$ \cite{bib-thrust}.  Events are
also required to be in that region of the detector with good \dEdx\ 
and silicon micro-vertex coverage by requiring
$\left|\cos(\theta_{T})\right|<0.75$, where $\theta_{T}$ is the polar
angle of the thrust axis.\footnote{%
  The right-handed OPAL coordinate system is defined such that the
  origin is at the center of the detector, the $z$-axis follows the
  electron beam direction and the $+y$ direction points up.  The polar
  angle~$\theta$~is defined relative to the $+z$-axis, and the
  azimuthal angle~$\phi$~is defined
  relative to the $+x$-axis.}  %
After the angular acceptance cut, thrust cut, and requiring important
detector components to be operational, $\mathrm{N_{mh}}$ = \NumTkmhAllCuts\ 
\mhic\ events are retained.

Reconstructed vertices displaced from the interaction point are used
to select \Ztobb\ events.  The primary vertex is determined for each
event using the average beam spot position as a constraint
\cite{\TauLifeUpdated,\TauLife}. Jets are found using a cone algorithm
with a cone having a half-angle of 0.55 radians and a minimum jet
energy of 5.0 GeV \cite{\OPALcone}.  Both charged tracks and
calorimeter clusters not associated with a track are used to identify
jets.  An iterative approach is used when attempting to form a
significantly displaced vertex, referred to as a secondary vertex, in
each jet \cite{\OPALbstars}.  Events containing at least one
reconstructed secondary vertex are retained.  The efficiency of
tagging a jet associated with a b quark is measured to be $(21.2\pm
0.9)\%$ for a purity of $(95.5\pm 0.5)\%$.  Details of the efficiency
calculation are described in section~\ref{sec:btag}.


\section{\symtitle{\Lamb} Identification} \label{sec:lambid}

Events are split into hemispheres using the plane orthogonal to the
thrust axis.  \Lamb\ particles are identified both in the hemisphere
with a jet containing a secondary vertex and in the opposite
hemisphere to increase the sample size.

The \Lamb\ selection used here is similar to the method described in
\cite{\OPALlambdaID}.  \Lamb's are reconstructed via the decay
\LambPPi.  All combinations of well--measured oppositely-charged
tracks forming a vertex are considered.  The higher momentum track is
assumed to be the proton.  Each track is required to have a
significant impact parameter in the r-$\phi$ plane with respect to the
primary vertex to reduce combinatorial backgrounds.  The \Lamb\ 
direction is required to be in the range $|\cos(\theta)_{\Lambda}| <
0.9$.  The momentum component parallel to the beam line for each
track, \pz, is re-calculated assuming it originates from the
reconstructed \Lamb\ decay point \cite{\OPALblamleppbr}.  \Lamb\ 
candidates whose invariant mass lies within 8 \MeVcc\ of the nominal
\Lamb\ mass are accepted if the invariant mass, assuming pion masses
for both tracks, does not fall within 6 \MeVcc\ of the
$\mathrm{K_{s}^{0}}$ mass.

The \dEdx\ identification of the candidate proton requires that the
observed energy loss is consistent with a proton and inconsistent with
that of a pion of the same momentum.  No \dEdx\ requirements are made
on the pion candidate.

The reconstructed decay point of the \Lamb\ is required to be at a
distance greater than 8 cm in the r-$\phi$ plane from the primary
interaction point.  There must be no hits in the silicon detector that
are associated with either of the tracks.  The angle $\phi(\Lamb)$
between the position vector of the \Lamb\ decay vertex and its
momentum vector is required to be less than 14 mrad.  To reduce
\Lamb's coming from fragmentation, the opening angle of the \Lamb\ 
direction with the jet axis is required to be less than 0.2 radians,
and the momentum of the \Lamb\ is required to be greater than
\Pmincut~\GeVc.
 
Monte Carlo studies indicate that the fake \Lamb 's remaining in
\Ztobb\ events are predominantly from real \Lamb\ decays where one
decay product of the \Lamb\ is combined with a random track.  The fake
\Lamb\ rate has been studied using side-bands of the \Lamb\ mass
distribution and, for the selection criteria of this paper, is
estimated at 2\% in both data and \mc.  The above selection retains
\NumLambDvCut\ events.


\section{Backgrounds} \label{sec:backgrounds}

Besides the \BlamLambX\ signal, the selected event sample contains the
following backgrounds: (1) events with a \bhadron\ and a \Lamb\ 
produced in the hadronisation process, (2) \Lamb's from \bmeson\ 
decay, (3) other backgrounds and fake \Lamb\ baryons.  These
three background sources are discussed below.\\

\noindent
(1) Events where a \Lamb\ baryon arises in hadronisation can be
separated from those produced in \Blam\ decays on a statistical basis.
The \Lamb\ baryons from this source generally have low momentum, \mom,
and transverse momentum, \Pt.\footnote{The transverse momentum of the
  \Lamb\ is measured with respect to the nearest jet axis.  The \Lamb\ 
  is included in the calculation of the jet direction.}  There are two
distinct sub-classes to this background:

\begin{itemize}
\item the \Lamb\ is created in associated production with another
  light baryon within a \Ztobb\ event;
\item the \Lamb\ is created in associated production with a primary
  \bbaryon.
\end{itemize}

\mc\ studies indicate that these two classes of \Lamb\ baryons
have similar \mom\ and \Pt\ distributions, which are different from
those of the signal.  Before the minimum momentum cut, the momentum
spectrum for fragmentation \Lamb 's peaks at 1~GeV, and \Pt\ is peaked
at 200~MeV.  Because these distributions have long tails, it is not
possible to remove all fragmentation
\Lamb 's with a cut.\\

\noindent 
(2) \BLambX\ decays are a significant background.  Although the
branching ratio for this process is small, the fraction of b quarks
which hadronise to \bmeson s is \btobmeson\ in \Zzero\ decays
\cite{\PDG}, so the \Lamb\ yield from \bmeson s is about the same as
that from \Blam\ baryons.  However, \Lamb's from \bmeson\ decays have
a softer \Pt\ spectrum than the signal since baryon number
conservation requires an additional baryon in the decay products of
the \bmeson.

While not yet observed, it is expected that the \Bs\ meson will
produce \Lamb's in its decay chain.  The kinematics are assumed to be
similar to \Bd\ decays.  Excited \bmeson s are also produced in
\Ztobb\ hadronisation.  We assume that these mesons decay either
hadronically or electromagnetically to a weakly decaying \bmeson, and
that the kinematics are similar to when the \bmeson s are directly
produced in
the ground state.\\

\noindent 
(3) Other backgrounds contribute 3\% to the sample.  The \Dplus\ is
the only charmed hadron with a lifetime long enough to produce a
signficant background after requiring a displaced vertex.  However,
since the \Dplus\ is too light to decay to a \Lamb\ and an
anti--baryon, it is only a background when coupled with a
fragmentation \Lamb\ or a charm baryon to \Lamb\ decay in the opposite
hemisphere.  The few \Dplus\ events accepted are included in the
background class (1) above, since they are kinematically similar.
Leading \Lamb\ baryons from light quark (u,d,s) decays of the
\Zzero\ are less than 1\% of the sample due to the secondary vertex
requirement.


\section{Fitting the \symtitle{\Lamb}\ Momentum and Transverse
  Momentum Spectra} \label{sec:fit}

\begin{figure}[tb]
  \begin{center}
    \epsfig{file=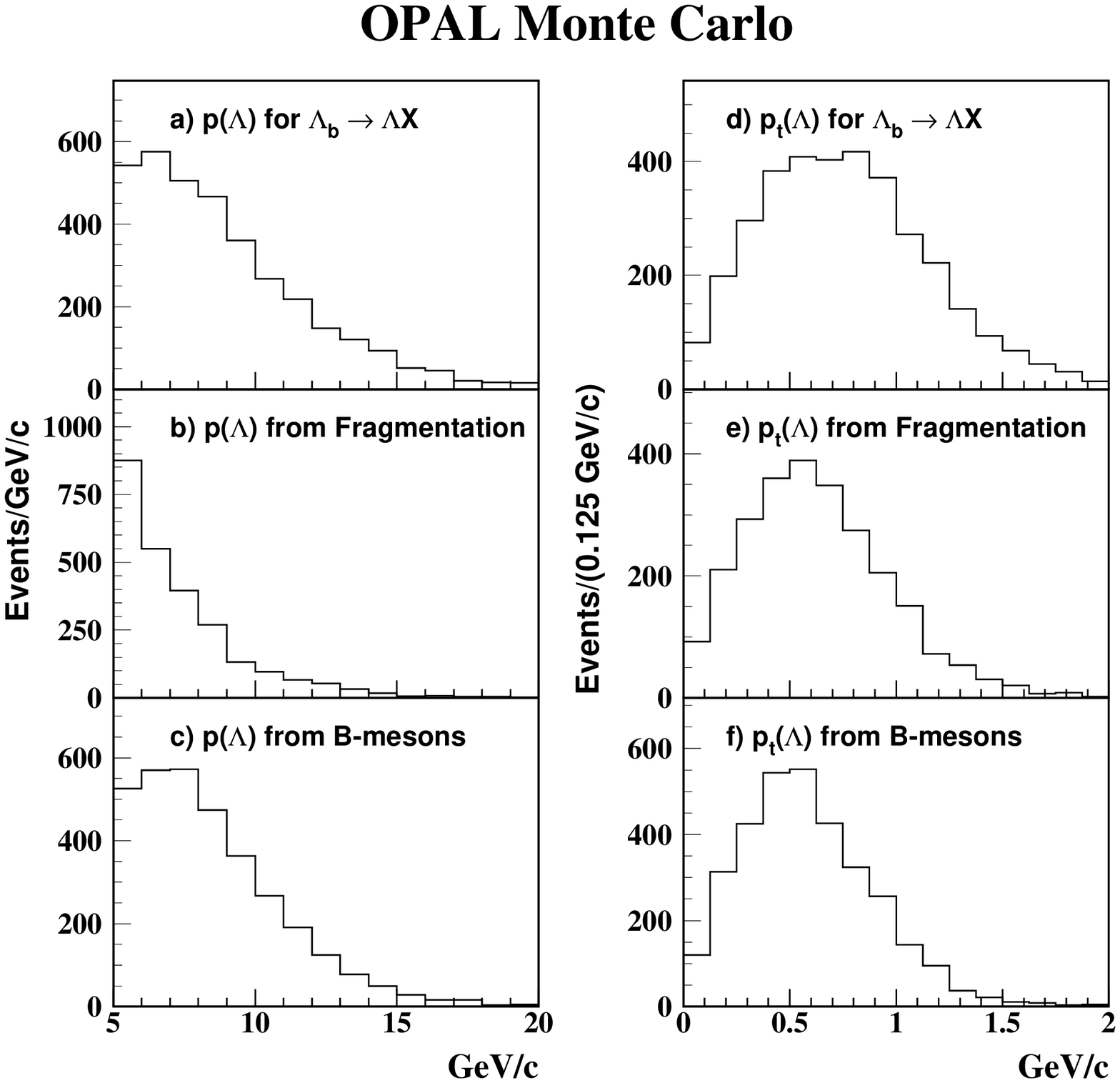,width=1.0\textwidth}
   \caption[\mc\ distributions for \protect\mom\ and \protect\Pt]
   {\label{fig-six-plots-mc-ppt} The \mc\ \mom\ and \Pt\ distributions
     for the signal and the two main background sources of \bhadron\ 
     events containing \Lamb 's.  The six distributions are used as
     input to the fit.}
  \end{center}
\end{figure}

The fraction of \Lamb 's from \BlamLambX\ decays is determined by
simultaneously fitting the total and transverse momentum spectra of
the \Lamb\ particles.  Figure \ref{fig-six-plots-mc-ppt} shows the six
input \mc\ distributions used in the fit.  They are the \mom\ and \Pt\ 
for \Lamb's from \bbaryon s and the two backgrounds: \bmeson\ and
fragmentation. (The uncertainties due to these \mc\ distributions are
discussed in section~\ref{sec:sysmc}.)  The fit returns the fractions
of the signal and the two major backgrounds in the data.

The fit, which takes into account finite statistics, uses a binned
maximum likelihood technique described in \cite{\BarlowBeeston}.  The
25\% correlation between \mom\ and \Pt\ is not taken into account by
this technique.  The effect of correlations was studied using \mc\ 
samples.  This technique underestimates the error by 5\%, but the
central value of the fit fraction is unchanged.  We correct the error
for this effect.

\begin{figure}[tb]
  \epsfig{file=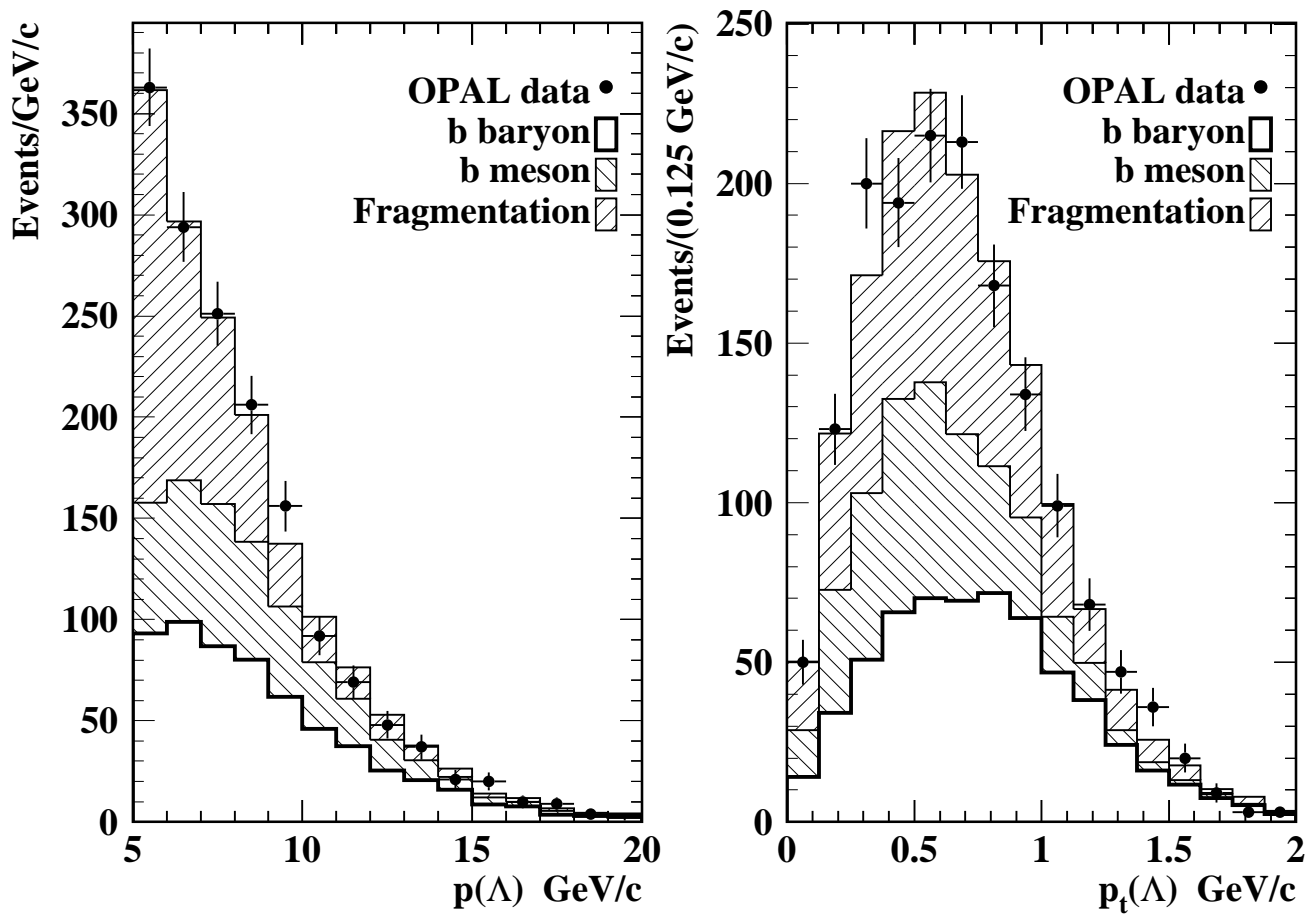,width=\textwidth
           ,bbllx=20pt,bblly=15pt,bburx=521pt,bbury=378pt}
   \caption[Fit fractions to \protect\mom\ and \protect\Pt\ from each source] 
   {\label{fig-PandPtfit} Distributions of \Lamb\ \mom\ and \Pt\ in
     data (points with error bars) and \mc\ (histogram).  The three
     sources: fragmentation, \bmeson, and \Blam\ are normalized to
     their fitted fractions.}
\end{figure}

The fitted distributions are shown in Figure~\ref{fig-PandPtfit} and
the fractions are listed in Table \ref{tab-FitResults}.  The $\chi^2$
of the fit is 25 for 27 degrees of freedom.  The fraction of signal
events is fitted to be \BlamPercent, where the error is due to finite
statistics in data and Monte Carlo.  This corresponds to \NumBlamCorr\ 
\BlamLambX\ signal events.

The fit method was checked by performing 5000 trial fits on fake data
samples which were generated by adding the three \mc\ sources and
allowing the histogram bins to vary according to Poisson statistics.
The distribution of fit fractions matches the values in the fake data
with a standard deviation equal to the uncertainty assigned by the
fitting routine.

The stability of the fit result was checked by varying the minimum
\mom\ and \Pt\ allowed in the fit and recalculating the product
branching ratio for each case.  The \Lamb\ momentum was allowed to
vary from 3 \GeVc\ to 8 \GeVc\ in 1 \GeVc\ steps.  We also varied the
minimum \Pt\ cut from 0.0 to 0.75 \GeVc\ in 0.25 \GeVc\ steps while
holding the minimum \mom\ cut constant at 5 \GeVc.  For all cases the
product branching ratio was consistent within $10\%$ and there were no
deviations that were not compatible with statistical fluctuations.

\begin{table}[tb]
\begin{center}
\begin{tabular}{|l|c||c|c|c|} \hline
Source & Fit fraction (\%) & \multicolumn{3}{c|}{Correlations}\\\cline{3-5}
 & & \BlamLambX & \FragLambX & \BLambX \\\hline
\BlamLambX & 37.4 $\pm$ 5.3 & 1.0   & --0.15 & --0.64 \\
\FragLambX & 37.1 $\pm$ 5.1 & --0.15 & 1.0   & --0.61 \\
\BLambX & 25.5 $\pm$ 6.6    & --0.64 & --0.61 & 1.0 \\\hline
\end{tabular}
\end{center}
\caption{\label{tab-FitResults} Results of the fit.  The fit fractions
  indicate the fraction of each source in the data sample.
  The correlations between the fit fractions are also shown.}
\end{table}


\section{Secondary Vertex Reconstruction Efficiency} \label{sec:btag}

The overall secondary vertex reconstruction efficiency, \epsdv,
depends on the efficiency of reconstructing a secondary vertex in both
the unbiased b hemisphere, \epsb, and the hemisphere containing the
decay \BlamLambX, \epsBlamLambX.  To measure the efficiency of
reconstructing a secondary vertex in an unbiased \bhadron\ hemisphere,
we compare the fraction of tagged hemispheres in events with at least
one reconstructed secondary vertex to the number of events with a
reconstructed vertex in both hemispheres.  This is done for a sample
of \mh\ events which pass all selection criteria except \Lamb\ 
identification.

Solving the following equations for \epsb\ yields the efficiency.
\begin{equation}
  \Rb\epsb + \Rudsc\epudsc = f_\mathrm{1v}
\end{equation}
\begin{equation}
  \Rb\epsb^2 + \Rudsc\epudsc^2 = f_{\mathrm{2v}}
\end{equation}
The efficiency of selecting a non-b hemisphere is represented by the
symbol \epudsc, $f_\mathrm{1v}$ is the fraction of hemispheres with a
reconstructed secondary vertex, $f_{\mathrm{2v}}$ is the fraction of
events in which both hemispheres have a secondary vertex, \Rb\ is
$\Gambb/\Gamhad=0.2169$ \cite{\PDG} and $\Rudsc\ = 1-\Rb$.  This
measurement yields a value for \epsb\ of $(21.2 \pm 0.9$)\%, where the
error is from finite statistics in the data and \mc, and year-by-year
variations in the detector configuration.  The effect of tagging
efficiency correlations between the hemispheres is negligible for
this analysis. 

The presence of a high-momentum \Lamb\ in a hemisphere reduces the
efficiency for reconstructing a secondary vertex.  This is true for
all sources of \Lamb's in the sample.  The proton and pion from
\Lamb's in signal events are unlikely to be included in a secondary
vertex, which reduces its probability of being reconstructed.  The
\Lamb\ selection enhances the proportion of \Blam 's which have a
shorter lifetime than the average \bhadron.  \mc\ studies indicate
that, when the average \bhadron\ lifetime is adjusted to that of the
\Blam, the shorter lifetime reduces the secondary vertex
reconstruction efficiency by a factor of $0.84 \pm 0.05$.  Lastly, if
the selected high momentum \Lamb\ is from fragmentation, the primary
\bhadron\ will have less momentum than usual.  A shorter flight
distance decreases the probability of reconstructing a displaced
vertex.

To calculate the efficiency for reconstructing secondary vertices in
\BlamLambX\ hemispheres, we begin by comparing the number of selected
\Lamb 's in the same hemisphere as a reconstructed secondary vertex
(same-side) to the number of \Lamb's in the opposite hemisphere
(opposite-side).  If the presence of high-momentum \Lamb's had no
effect on the reconstruction of vertices we would expect to find the
same number of same-side and opposite-side \Lamb's in the sample.
Instead, we find that the ratio of same-side to opposite-side \Lamb's
is $R^{(data)}=0.58 \pm 0.03$ and $R^{(MC)}=0.64 \pm 0.01$.  In the
\mc\ the ratios for specific sources of \Lamb's are:
$R_{\Blam}^{(MC)}=0.54 \pm 0.03$, $R_{frag}^{(MC)}=0.62 \pm 0.03$,
$R_{\bmeson}^{(MC)}=0.77 \pm 0.03$.

$R^{(MC)}$ must be multiplied by a factor of 0.9 to match
$R^{(data)}$.  Assuming that this factor is the same for each source,
the corrected ratio for \Blam\ is $R'^{(data)}_{\Blam} = 0.49 \pm
0.06$, where the full size of the correction is included in the
uncertainty.  The efficiency for tagging hemispheres with the decay
\BlamLambX\ is

\begin{equation}
  \epsilon_{\BlamLambX}^{b} = \epsb\cdot R'^{(data)}_{\Blam} = (10.4 \pm
1.3) \%
\end{equation}
where the error includes both statistical and systematic
uncertainties.

An event containing a \BlamLambX\ decay may be tagged by a
reconstructed secondary vertex either in the hemisphere containing the
\Lamb\ or in the opposite hemisphere.  The overall efficiency for
identifying displaced vertices in these events is therefore:
\begin{equation}
  \epsdv=\epsilon_{\BlamLambX}^{b}+
                \epsb(1-\epsilon_{\BlamLambX}^{b})=(29.4\pm1.5)\%.
\end{equation}


\section{\symtitle{\Lamb} Reconstruction Efficiency} \label{sec:lambeff}

The efficiency of \Lamb\ reconstruction is determined from \mc\ for
\Lamb 's satisfying all selection criteria. The \mc\ simulates well
the kinematic properties like \Lamb\ mass resolution, momentum
distributions, and the \dEdx\ response of the detector.  The overall
efficiency for reconstructing \Lamb\ particles from \Blam\ decay,
\epsLam, is found to be \Lambeff\ for a minimum momentum of
\Pmincut~\GeVc.  The error comes from \mc\ statistics, the 2\% fake
rate and the tracking resolution.  The sensitivity to tracking
resolution was studied by varying the \mc\ resolutions by $\pm 10\%$
which caused the \Lamb\ identification efficiency to change by
$\pm0.4\%$.


\section{Systematic Uncertainty due to \mc\ Distributions} \label{sec:sysmc}

The simultaneous fit to the momentum spectra of the \Lamb\ candidates
requires six \mc\ input distributions: the \mom\ and \Pt\ of \Lamb 's
from \bbaryon, \bmeson, and fragmentation sources.  A systematic error
is assigned to account for possible mis-modelling of these
distributions.  Each distribution is checked against a data sample,
though the comparisons are limited because it is impossible to obtain
pure, large data samples for the three sources.

This section describes how the uncertainties on each of the \mc\ 
distributions are determined and how they propagate to the fit
fraction for the \Blam\ source, $f_{\Blam}$.  The results are
summarised in Table \ref{tab-fBlam}.

\begin{table}
\begin{center}
\begin{tabular}{|l|c|c|}
\hline
Sources of Systematic Errors for $f_{\Blam}$ & negative & positive \\
 & errors & errors \\\hline
\mom(\Lamb) and \Pt(\Lamb) from \BLambX\ & --3.5\% &
3.5\% \\
\mom(\Lamb) of Fragmentation \Lamb's &
--1.3\% & 0.8\% \\
\Pt(\Lamb) of Fragmentation &
--12.5\% & 12.7\% \\
\mom(\Lamb) from \BlamLambX\ & --2.9\% & 4.8\% \\
\Pt(\Lamb) from \BlamLambX\ & --16.6\% & 19.8\% \\
Tracking Uncertainty & --2.7\% & 2.7\% \\\hline
Total                                & --21.5\%   & 24.4\% \\\hline
\end{tabular}
\end{center}
\caption{\label{tab-fBlam} Systematic errors contributing to
the uncertainty in the measurement of the \Blam\ fit fraction, $f_{\Blam}$.}
\end{table}

\subsection*{\symtitle{\Lamb}'s from \bmeson s}

The \mc\ momentum spectra of \Clam 's and \Lamb 's coming from \bmeson
s are adjusted to match CLEO data \cite{\CLEOBLamb}.  In the \bmeson\ 
rest frame, the CLEO data have large errors for \Lamb 's with momentum
less than 0.5 GeV/c.  \Lamb's with low momentum in the rest frame of
the \bmeson\ are reweighted within a range corresponding to these
uncertainties, and the fit is repeated for several reweightings in the
selected range, changing the \Blam\ fit fraction by 7\%.  Values in
the center of the range are used in the final fit, and an error of
3.5\%, half of the variation observed, is assigned for the uncertainty
in the B meson \mom\ and \Pt\ distributions.

\subsection*{\symtitle{\Lamb}'s from Fragmentation}

A powerful technique for isolating \Lamb 's coming from \Blam\ baryons
requires that a lepton with high momentum and transverse momentum be
identified in the hemisphere with the \Lamb\ 
\cite{\OPALlamlep,\OPALblamlife}.  For these studies lepton refers to
only electrons and muons.  The correlation between lepton charge and
baryon number of the \Lamb\ is indicative of its origin: combinations
with opposite lepton charge and baryon number (right-sign) are used to
tag \BlamLambLepX\ events; wrong-sign combinations yield a high purity
of fragmentation \Lamb's \cite{\OPALPGagnon}, which are used as a
control sample to compare the \mom\ and \Pt\ spectra of fragmentation
\Lamb's in the data and Monte Carlo.

The lepton identification of \cite{\OPALleptonID} is used.  The \Lamb\ 
selection is tuned to maximize the number and purity of \Lamb 's from
fragmentation in the wrong-sign sample.  Differences with respect to
the event selection of section~\ref{sec:lambid} include removing the
secondary vertex requirement and requiring the invariant mass of the
\Lamb--lepton pair to be greater than 2 \GeVcc\ to reject \Clam\ and
\bmeson\ decays.  With this selection, 266 data events are found with
a purity of fragmentation \Lamb's of 75\%.

\begin{figure}[tb]
  \epsfig{file=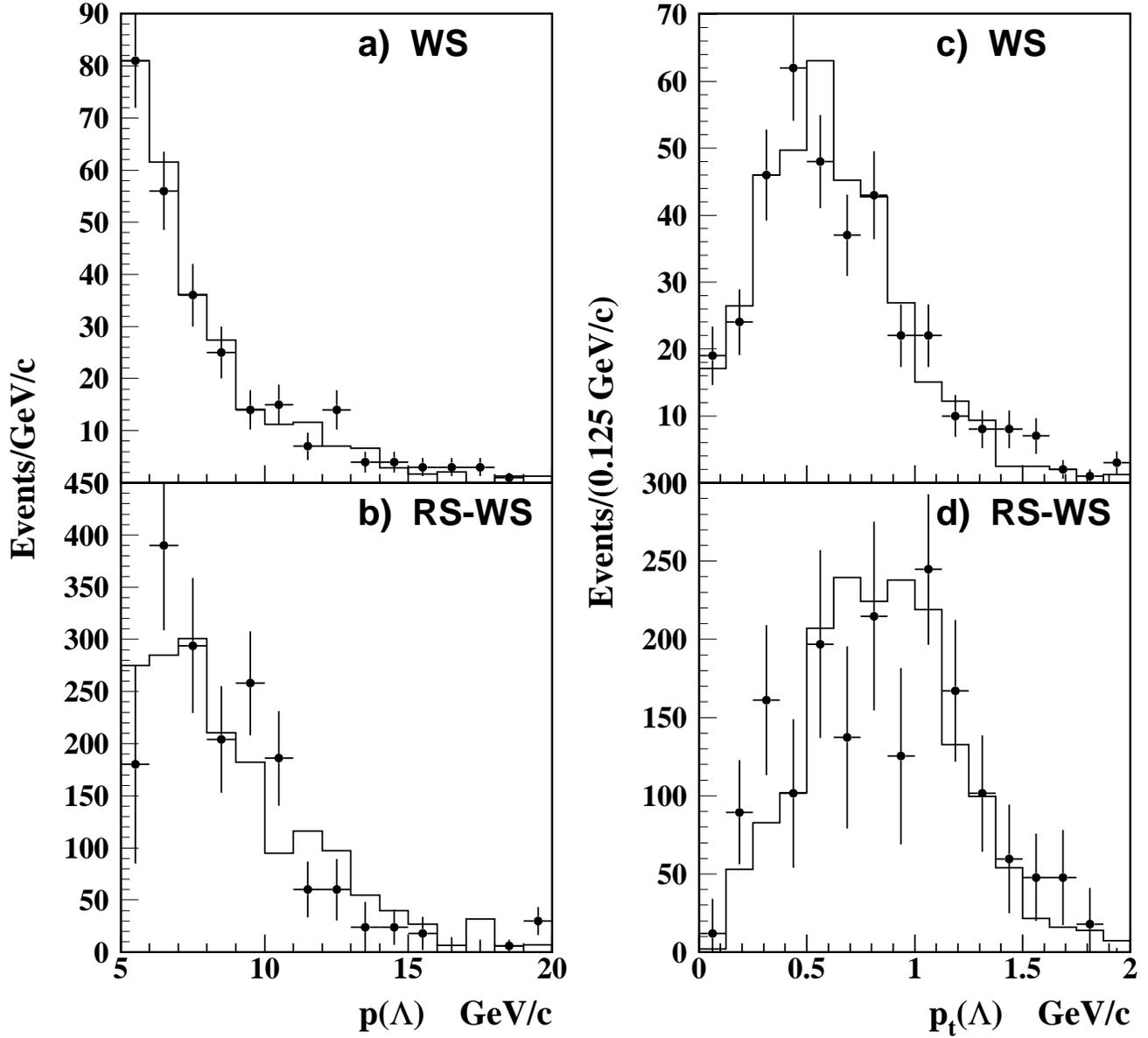,width=\textwidth
           ,bbllx=10pt,bblly=8pt,bburx=553pt,bbury=522pt}
   \caption[Wrong-sign and Right-sign minus Wrong-Sign distributions 
   from \Lamb--lepton studies.]  {\label{fig-lamlep-bothwsrs} Plots a)
     and c) represent wrong-sign distributions from \Lamb--lepton
     studies used to compare \Lamb's from fragmentation in data and
     \mc.  Plots b) and d) represent the results of the right-sign
     minus wrong-sign subtraction used to compare \Lamb's from
     \Blam's.  Points with error bars are OPAL data and the histogram
     is \mc.}
\end{figure}

Plots a) and c) of Figure \ref{fig-lamlep-bothwsrs} show the
comparison of the data and \mc\ wrong-sign distributions.  The means
of the \mom\ and \Pt\ distributions for data and \mc\ agree well and
are listed in Table \ref{tab-RSWS-errors}.  We calculate an
uncertainty in the agreement by adding in quadrature the errors on the
data and \mc\ means.  An uncertainty of 2.4\% is assigned for the
momentum distribution and 4.2\% for the transverse momentum
distribution.

To assess a systematic error on $f_{\Blam}$ due to the \mc\ simulation
of \mom\ or \Pt, each entry in a distribution is multiplied by a
factor which raises or lowers the mean of the distribution.  For
example, to see the effect of the 2.4\% uncertainty on the \mom\ mean
we multiply each entry by 0.976, which shifts the mean downward.  The
fit is then repeated with the shifted distribution to observe the
change in $f_{\Blam}$.  The procedure is repeated with a factor of
1.024 to determine the positive error on $f_{\Blam}$.

The fit fraction $f_{\Blam}$ changes by $^{+0.8}_{-1.3}$ \% when the
fragmentation p distribution is varied by its uncertainty, and by
$^{+12.7}_{-12.5}$ \% when the \Pt\ distrubution is varied. These
changes are assigned as systematic errors on $f_{\Blam}$ due to
uncertainties in the fragmentation \Lamb\ spectra.

\begin{table}
\begin{center}
\begin{tabular}{|l||c|c|c|c|}
\hline
& $\mu_{data}$ (\GeVc) & $\mu_{MC}$ (\GeVc) & $\delta\mu$ & $\sigma_{\mu}$ \\\hline

WS \mom(\Lamb) & 7.70 $\pm$ 0.15 & 7.55 $\pm$ 0.09 & 2.0\% &  2.4\% \\
WS \Pt(\Lamb)  & 0.70 $\pm$ 0.03 & 0.67 $\pm$ 0.01 & 4.0\% &  4.2\% \\

RS-WS \mom(\Lamb) & 8.63 $\pm$ 0.28  &   8.84 $\pm$ 0.10& 2.4\% & 3.4\% \\
RS-WS \Pt(\Lamb)  & 0.85 $\pm$ 0.04 &  0.86 $\pm$ 0.01 & 1.2\% & 5.2\% \\

\hline
\end{tabular}
\end{center}
\caption{\label{tab-RSWS-errors}  A summary of the \Lamb-lepton 
  studies comparing data and \mc\ for right-sign minus wrong-sign 
  (RS--WS) and wrong-sign (WS) distributions.  RS-WS is used to
  compare \Lamb\ \mom\ and \Pt\ for \Lamb's in \BlamLambLepX\ decay.
  WS is used to
  compare distributions from fragmentation \Lamb's.  The means of the
  \mc\ and data distributions are listed along with their relative
  difference $\delta\mu$ and the uncertainty on the relative
  difference $\sigma_{\mu}$ which is from the data and \mc\
  uncertainties added in quadrature.}
\end{table}

\subsection*{\symtitle{\Lamb}'s from \symtitle{\Blam}}

The technique described in the previous section is also used to
compare \mom\ and \Pt\ distributions for \Lamb's from \BlamLambLepX\ 
decays.  Right-sign combinations of \Lamb's and leptons result in a
sample composed mostly of signal \Lamb's, with the remainder being
\Lamb's from fragmentation.  To isolate the signal shapes, the
wrong-sign distributions are subtracted from the right-sign.
Fragmentation \Lamb's and those from light quark events populate the
right-sign and wrong-sign equally, so the background subtracted
distributions represent well the momentum distributions of \Lamb's
from \BlamLambLepX\ decay.

The event selection used here is the same as for the previous section,
except that the minimum \Lamb-lepton invariant mass cut is set at 2.2
\GeVcc, which reduces contributions from \bmeson\ and \Clam\ decays to
about one percent of the total.  After subtracting wrong-sign from
right-sign, the momentum distributions in data contain 289 entries.
The \mom\ and \Pt\ of the subtracted distributions agree well.  Values
are listed in Table \ref{tab-RSWS-errors} and the distributions are
shown in plots b) and d) of Figure \ref{fig-lamlep-bothwsrs}.

Shifting the \BlamLambX\ \mc\ distributions by $\pm$3.4\% for \mom\ 
and $\pm$5.2\% for \Pt\ yields systematic uncertainties on $f_{\Blam}$
of $^{+4.8}_{-2.9}$ \% from the \mom\ distribution and
$^{+19.8}_{-16.6}$\% from \Pt.

In addition to these tests, we also varied the shape of the \Pt\ 
distribution to ensure that shifting the means was a good measure of
how variations affect the fit.  For several reweightings which skewed
the distribution to look more like the semileptonic distribution, it
was found that the fit result was directly correlated to the mean \Pt.
Variations in the mean are a good measure of the sensitivity of the
fit to variations in the input \mc\ distributions.


\section{Other Systematic Uncertainties} \label{sec:othersys}

This section discusses systematic uncertainties that have not already
been addressed.  They are the effect of \Blam\ polarisation
\cite{\OPALblampol}, \mc\ decay model, and tracking resolution.

\Blam\ polarisation was not simulated in the \mc.  The presence of
\Blam\ polarisation shifts slightly the momentum distributions of
\Lamb 's coming from \Blam 's.  Possible differences between \mc\ and
data due to polarisation are, therefore, included in the previously
assigned uncertainties because all \mc\ distributions are compared
directly to data samples.

Three additional factors which govern the primary hadronisation
process and decay of the \Blam\ and its daughter particles could each
affect the \Lamb\ momentum distributions.  These are the modelling
of b quark fragmentation \cite{\PetersonFrag} and baryon production
and decay in the \mc.  Again, since the \Lamb\ \mom\ and \Pt\ for all
sources have been compared directly to data samples, the effects are
small and any contributing uncertainties are included in errors
assigned for the modelling of the momentum distributions.

The effect of tracking resolution is also small for this analysis.
Varying the \mc\ tracking resolutions by $\pm 10\%$, consistent with
the known quality of tracking in the simulation, causes the \Blam\ fit
fraction, $f_{\Blam}$, to vary by $\pm 0.01$.  This is included in the
uncertainty of $f_{\Blam}$ in Table \ref{tab-fBlam}.


\section{Consistency Checks of \symtitle{\Pt} for \symtitle{\Lamb}'s
  from \symtitle{\Blam}} \label{sec:checks}

Since the overall systematic error is most sensitive to the \Lamb\ 
\Pt\ from \Blam\ decays, we present three additional checks on this
distribution.  While these tests are not as precise as the
\Lamb--lepton studies described in section~\ref{sec:sysmc}, they are
consistent and provide additional evidence that the data is
well modelled for \BlamLambX\ decays and that semi-leptonic decays
provide a good measure of the \mc\ uncertainty.\\

\noindent (1) One test of the modelling of \Lamb\ \Pt\ from \BlamLambX\ decays
is made using the fit directly. The \Pt\ distribution is shifted until
the \chisq\ of the fit exceeds the 90\% confidence interval.  This
occurs for shifts of +6\% and --13\%.  These limits suggest that the
true mean of the \Pt\ distribution is somewhere in this range.  Hence
the possible increase in the mean \Pt\ for \Lamb's from \Blam\ decay
cannot be very much greater than the assigned uncertainty of 5.2\%.\\

\noindent (2) This test is similar to the \Lamb--lepton studies already described
in section~\ref{sec:sysmc}, but looks for a lepton in the opposite
hemisphere from the \Lamb.  This lepton does not bias the inclusive
sample of \Lamb's, and therefore provides a direct check of \Lamb's
from \BlamLambX\ decays.  When the b quark does not mix in the lepton
hemisphere, the correlation between charge and baryon number (with the
opposite definition of right-sign and wrong-sign) still holds for
\Blam\ decays.  Fragmentation \Lamb\ events still populate the
right-sign and wrong-sign samples equally.

Because an invariant mass cut between the \Lamb\ and lepton is no
longer meaningful, there are a significant number of \bmeson\ decays
in the right-sign sample. After subtracting the wrong-sign \Lamb\ \Pt\ 
distributions from the right-sign, the \mc\ predicts a sample with
$65\%$ \Blam\ events, and the rest from \bmeson s.  The average \Pt\ 
in data and \mc\ agree to within a statistical precision of 10\%.
This large uncertainty is due to the many fragmentation and \bmeson\ 
events in both the right and wrong-sign samples.  While not as
powerful as the studies of same-side semileptonic decays, the
agreement is
evidence of good \mc\ simulation of non-leptonic \Blam\ decays.\\

\noindent (3) Finally, the \mc\ predicts a \Lamb\ \Pt\ mean 13\% higher for
\BlamLambLepX\ decays than for \BlamLambX\ decays.  Three factors
contribute to this difference: a 10\% correlation between the \Pt\ of
the lepton and \Lamb, the \Lamb--lepton invariant mass cut, and a
smaller multiplicity in semileptonic \Blam\ decays.  Understanding
this difference offers us some confidence that \BlamLambX\ decays are
well modelled.

The effect of correlation between the lepton and \Lamb\ \Pt\ is
investigated by varying the minimum lepton \Pt\ cut and observing the
corresponding shift in the \Lamb\ mean \Pt.  For minimum lepton \Pt\ 
cuts between 0. and 1.5 \GeVc\ with 0. as the reference value, a
maximum variation of 5\% is seen in the \Lamb\ mean \Pt.  The \mc\ 
models well the effect in the data.  Similarly, the effect of the
\Lamb--lepton invariant mass cut is investigated by varying its value.
Once again the Monte Carlo models the effects on the data well, with a
7\% observed variation in the \Lamb\ mean \Pt\ in response to changes
in the invariant mass cut.  Together, the \Lamb--lepton correlation
and the invariant mass cut account for more than half of the 13\%
difference in the \Pt\ means for \Lamb's from \BlamLambLepX\ and
\BlamLambX\ decay.

The remaining difference in the mean \Pt\ values can be accounted for
by the different multiplicities of the two types of decay.
Fundamentally, the differences in \Lamb\ \Pt\ between semileptonic and
hadronic decays of the \Blam\ is governed by differences in decay
multiplicity.  The semileptonic decays are expected to have a lower
multiplicity, which will cause the mean \Pt\ to be
higher.\footnote{The \Lamb\ momentum in the lab frame is not as
  significantly affected by decay multiplicity.  It is dominated by
  boost effects.}  This effect is investigated by examining the track
multiplicity in secondary vertices that are in the same hemisphere as
the \Lamb.

In the \mc, the average multiplicity for vertices in hemispheres with
a \Lamb--lepton pair is lower than for hemispheres with just a
selected \Lamb\ by $17\pm 2\%$.  In data, this difference is $10\pm
5\%$. The \mc\ adequately models the multiplicity difference,
providing further evidence that the \Pt\ differences between
\BlamLambLepX\ and \BlamLambX\ events are well modelled


\section{Results}  \label{sec:results}

The product branching ratio can be expressed as follows:
\begin{equation}
  \PBRBlamLambX = \frac{N_{\Lamb} f_{\Blam}}{2 R_b \Nmh \epsLam \epsdv
    \BRLambPPi}
\end{equation}
where $N_{\Lamb}$ is the number of \Lamb\ candidates in the final
sample, and $f_{\Blam}$ is the fitted fraction of \Lamb 's from \Blam
's.  \Rb\ is the fraction of hadronic \Ztobb\ decays \cite{\PDG},
\Nmh\ is the number of identified \mhic\ events in the data, \epsLam\ 
is the efficiency of reconstructing \LambPPi\ decays from \Blam,
\epsdv\ is the overall secondary vertex reconstruction efficiency for
\BlamLambX\ events, and \BRLambPPi\ is the fraction of \Lamb 's
decaying to a proton and pion \cite{\PDG}.  See Table
\ref{tab-Results} for the values of these quantities.

\begin{table}[tb]
\begin{center}
\begin{tabular}{|l|c|c|} \hline
\hspace{5mm}  &   & Uncertainty on \\
\hspace{5mm}                   &        & \PBRBlamLambX \\\hline
\hspace{5mm}$N_{\Lamb}$    & $1582\pm40(stat)$  & $\pm 0.07(stat)$ \\
\hspace{5mm}$f_{\Blam}$    & $0.374\pm
0.053(stat)^{+0.091}_{-0.080}(sys)$ & $0.38(stat)^{+0.65}_{-0.57}(sys)$ \\
\hspace{5mm}\Rb            & $0.2169\pm 0.0012(sys)$  & $\pm 0.01(sys)$ \\ 
\hspace{5mm}\Nmh           & \NumTkmhAllCuts & --- \\
\hspace{5mm}\epsLam        & $0.117\pm 0.006(sys)$ & $\pm 0.14(sys)$ \\
\hspace{5mm}\epsdv         & $0.294\pm 0.015(sys)$ &  $\pm 0.14(sys)$ \\
\hspace{5mm}\BRLambPPi     & $0.639\pm 0.005(sys)$ &  $\pm 0.02(sys)$ \\\hline

\end{tabular}
\end{center}
\caption{\label{tab-Results} Quantities needed for evaluating the
  product branching ratio.}
\end{table}

Combining these factors, the measured product branching ratio is
\begin{equation}
  \PBRBlamLambX = \PBRAnswer.
\end{equation}
where the statistical error arises from the finite number of data and
\mc\ events in the fit and the systematic error is dominated by the
modelling of the \Lamb\ \Pt\ spectrum in \Blam\ decays.

To calculate a new OPAL value of the product branching ratio we
combine the result presented here with an updated value of the
previous OPAL measurement using the more recent value of
$\Rb=0.2169\pm 0.0012$.  This gives, for the previous product
branching ratio, $(4.00 \pm 0.47(stat) \pm 0.38(sys))\%$
\cite{\OPALPGagnon}.  The event samples for the two measurements of
the product branching ratio have less than 20\% of events in common.
Taking into account statistical and systematic correlations, we find
\begin{equation}
  \PBRBlamLambX = \OPALPBR.
\end{equation}
This agrees with and is of higher precision than the value of $(2.2
^{+1.3}_{-0.8})$\% measured by the DELPHI collaboration
\cite{\DELPHIpbr}.

The value of \fbBlam\ has been determined to be \btobbaryon\cite{\PDG}
using measurements of reconstructed $\Clam \ell$ and $\Xi \ell$
events.  Using this value, we calculate
\begin{equation}
  \mathrm{BR}(\BlamLambX) = \BRAnswer. 
\end{equation}
Finally, we use the data and method of \cite{\OPALPGagnon} to
calculate an improved value for the ratio $R_{\Lamb \ell} =
\RatBRBlamLambLepX = (8.0 \pm 1.2 \pm 0.9)\%$.

\clearpage
\medskip
\bigskip\bigskip\bigskip
\appendix
\par
{\bf Acknowledgements}
\par
\noindent We particularly wish to thank the SL Division for the efficient operation
of the LEP accelerator at all energies and for their continuing close
cooperation with our experimental group.  We thank our colleagues from
CEA, DAPNIA/SPP, CE-Saclay for their efforts over the years on the
time-of-flight and trigger systems which we continue to use.  In
addition to the support staff at our own
institutions we are pleased to acknowledge the  \\
Department of Energy, USA, \\
National Science Foundation, USA, \\
Particle Physics and Astronomy Research Council, UK, \\
Natural Sciences and Engineering Research Council, Canada, \\
Israel Science Foundation, administered by the Israel
Academy of Science and Humanities, \\
Minerva Gesellschaft, \\
Benoziyo Center for High Energy Physics,\\
Japanese Ministry of Education, Science and Culture (the Monbusho) and
a grant under the Monbusho International
Science Research Program,\\
Japanese Society for the Promotion of Science (JSPS),\\
German Israeli Bi-national Science Foundation (GIF), \\
Bundesministerium f\"ur Bildung, Wissenschaft,
Forschung und Technologie, Germany, \\
National Research Council of Canada, \\
Research Corporation, USA,\\
Hungarian Foundation for Scientific Research, OTKA T-016660,
T023793 and OTKA F-023259.\\

\clearpage

\bibliography{pr265} 

\end{document}